\documentclass[12pt]{iopart}

\usepackage{amsgen}
\usepackage{amssymb}
\usepackage{epsfig}
\usepackage{setstack}

\begin{document}

\title{Spherically Symmetric Solutions to Fourth-Order Theories of Gravity}

\author{T Clifton}

\address{DAMTP, Centre for Mathematical Sciences, University of
  Cambridge,  Wilberforce Road, Cambridge, CB3 0WA, UK.}

\eads{\mailto{\mailto{T.Clifton@damtp.cam.ac.uk}}}

\pacs{04.20.-q, 04.20.Jb, 04.25.Nx}

\begin{abstract}

Solutions to the field equations generated from Lagrangians of the form
$f(R)$ are considered. The spherically symmetric solutions to these equations
are discussed, paying particular attention to features that differ from
the standard Schwarzschild solution. The asymptotic form of solutions will
be described, as will the lack of validity of Birkhoff's theorem. Exact
solutions are presented which illustrate these points and the
stability and equations of motion of particles in these space-times
are investigated.

\end{abstract}

\maketitle

\section{Introduction}

It has been known for some time that Birkhoff's theorem and asymptotic
flatness are not ensured in the vacuum solutions of generalised fourth-order theories of
gravity, as they are in general relativity (GR) (see e.g. \cite{Pech,
  Ste77}).  The standard approach to overcoming this problem is to consider theories which contain
in their generating Lagrangian an Einstein-Hilbert term which
dominates the field equations in the low-curvature limit.  In this way
the usual behaviour of approaching Minkowski space at asymptotically
large distances from sources is assumed to occur and perturbative
expansions about a Minkowski background are then
possible.  The weak-field limit of the theory in question can then be
investigated in a straightforward way and compared with experiment.

Whilst the existence of a Minkowski background is a great
simplification, and very useful in investigations of the weak-field, it
is unclear whether or not a theory should be disregarded solely on the
grounds of this limit not existing.  There are a number of gravitational theories that can be
conceived of which do not always admit Minkowski space as a solution.  These
include theories derived from Lagrangians of the form $R^{1+\delta}$
(where $\delta \neq 0$) \cite{Cli, Bar06, further, Car04, Leach, Schmidt}, as well as those derived form
$R + \alpha/R^n$ (where $n > 0$) \cite{RRn, Car05}.  The latter
has caused considerable debate as to whether or not it is
compatible with solar system experiments.  Studies on this subject
usually follow the prescribed analysis for computing the weak
field geometry in GR; picking a highly symmetric background\footnote{often Minkowski
space, even though it is not always a solution of the field equations} and calculating
the form of spherically symmetric perturbations by linearising the
field equations and solving them, to first order in perturbations.
However, it is often unclear how an appropriate background to expand about
should be chosen, and how the lack of Birkhoff's theorem or asymptotic
flatness should be taken into account when performing these analyses.
We will discuss these points here and attempt to make some progress
into understanding the spherically symmetric situation by using
explicit exact vacuum solutions of the field equations.

In section 2 we present the field equations for the theories we will
be considering and two exact spherically symmetric solutions to these
field equations.  In section 3 we perform a linear perturbation
analysis about the backgrounds of the two exact solutions and in
section 4 we find the equations of motion for test-particles in these
space-times.  In section 5 we summarise our results.

\section{Field equations and exact solutions}

We will consider here gravitational theories derived from a Lagrangian of
the form
\begin{equation}
\label{L}
\mathcal{L} = f(R)
\end{equation}
where $f(R)$ is an arbitrary power-series expandable function of the
Ricci curvature scalar $R$.  Extremizing the action that is obtained
by integrating (\ref{L}) over all space then gives the field equations
\begin{equation}
f' R_{a b} -\frac{1}{2} f g_{a b} +{{f'}_;}^{c d}(g_{a b} g_{c d}-g_{a c}
g_{b d}) = 8 \pi G T_{a b}
\end{equation}
where matter sources have been included, primes denote
differentiation with respect to $R$ and we have disregarded boundary terms.  It can be seen that in the
low-curvature limit, $R << 1$ (assuming this limit exists), the term in the power-series
expansion of $f(R)$ with the lowest power will be the one which
dominates the field equations.  For this reason we specialise our
considerations, from this point on, to the choice $f(R)=R^{1+\delta}$
which reduces to general relativity in the limit $\delta \rightarrow
0$.  This choice is the theory considered in \cite{Cli, Bar06,
  further} and will be the effective theory at low curvatures for
$f(R)=R+\alpha/R^n$ when $\delta=-(1+n)$.  The field equations are
then given by
\begin{eqnarray}
\label{fieldq}
\fl \qquad \qquad \delta (1-\delta ^{2})R^{\delta }
\frac{R_{,a}R_{,b}}{R^{2}}&-\delta (1+\delta
)R^{\delta }\frac{R_{;ab}}{R}+(1+\delta )R^{\delta }R_{ab}-\frac{1}{2}
g_{ab}RR^{\delta } \\ &-
          g_{ab}\delta (1-\delta ^{2})R^{\delta }\frac{R_{,c}R_{,}^{\ c}}{R^{2}}
+\delta (1+\delta )g_{ab}R^{\delta }\frac{\Box R}{R} = 8 \pi T_{a b}. \nonumber
\end{eqnarray}
We will now present two exact spherically symmetric vacuum solutions to this
set of equations.  The first of these was found previously in
\cite{Cli} and is given by the line-element
\begin{equation}
\fl {\bf{Solution \; 1}} \qquad \qquad \qquad
ds^{2}=-A_1(r)dt^{2}+\frac{dr^{2}}{B_1(r)}+r^{2} d\Omega^2 \label{Chan}
\end{equation}
where 
\begin{eqnarray}
\nonumber
A_1(r)& =r^{2\delta \frac{(1+2\delta )}{(1-\delta )}}+\frac{C_1}{r^{\frac{
(1-4\delta )}{(1-\delta )}}} \\
B_1(r)& =\frac{(1-\delta )^{2}}{(1-2\delta +4\delta ^{2})(1-2\delta (1+\delta
))}\left( 1+\frac{C_1}{r^{\frac{(1-2\delta +4\delta ^{2})}{(1-\delta )}}}
\right) \nonumber
\end{eqnarray}
and $C_1$ is a constant.  This solution is conformally related to the $Q=0$ limit
of a solution found by Chan, Horne and Mann \cite{Cha95} and reduces to the
Schwarzschild solution in the limit $\delta \rightarrow 0$.  The second
solution is given by
\begin{equation}
\fl {\bf{Solution \; 2}}  \qquad \qquad ds^2= -A_2(r) dt^2 +a^2(t) B_2(r) (dr^2 +r^2 d
\Omega^2) \label{Fon}
\end{equation}
where
\begin{eqnarray}
\nonumber
A_2(r) = \left(
\frac{1-\frac{C_2}{r}}{1+\frac{C_2}{r}}\right)^{\frac{2}{q}}& 
\qquad a(t) = t^{\delta \frac{(1+2 \delta)}{(1-\delta)}}\\
B_2(r) = \left(1+\frac{C_2}{r}\right)^4 A(r)^{q+2 \delta-1}&
\qquad q^2 = 1-2 \delta +4\delta^2 \nonumber
\end{eqnarray}
and $C_2$ is a constant.  This solution is conformally related to one
found by Fonarev \cite{Fonarev} and again reduces to the Schwarzschild
solution in the limit $\delta \rightarrow 0$.

These two solutions can be seen to exhibit features not present in the
Schwarzschild solution of GR.  Both of these solutions are strongly
curved, but each displays this curvature in a different way.  Solution
1 is static and does not reduce to an $r$ independent form in the
limit $r \rightarrow \infty$ (despite the Ricci scalar approaching $0$
in this limit).  Solution 2 displays more conventional behaviour in
the limit $r \rightarrow \infty$, but shows explicitly the lack of validity
of Birkhoff's theorem.  This solution becomes $r$ independent in the
limit $r \rightarrow \infty$, but still displays strong curvature in
this limit as the metric reduces to the spatially flat vacuum
Friedmann-Robertson-Walker cosmology found in \cite{Cli}.

We will now continue to find the general form of spherically symmetric
perturbations to the backgrounds ($r \rightarrow \infty$ limit) of the
two exact solutions above.  It will be seen that there exist extra
modes which are not excited in the exact solutions, but that the modes
corresponding to the linearised exact solutions above are the ones most
interesting for performing gravitational experiments in these
space-times.

\section{Linear perturbation analysis}

Perturbative analyses in the literature are often performed about
Minkowski space or de-Sitter space.  This is
perfectly acceptable practise in GR and fourth-order theories in which
an Einstein-Hilbert term dominates in the low curvature regime.  In
other fourth-order theories, of the type considered here, in which the
Einstein-Hilbert term does not dominate the low curvature regime then
there is good reason to consider perturbing about other backgrounds.
We have shown explicitly, with exact solutions, the existence of other
spherically symmetric space-times.  We will now proceed to perform a
linear perturbation analysis about the backgrounds ($r \rightarrow
\infty$ limit) of these two space-times.  The general solution to
first order in perturbations will be found and it will be shown that
the form of these linearised solutions will be strongly dependant on
the background.

\subsection{Solution 1}

An analysis of the linear perturbations about the background
prescribed by equation (\ref{Chan}) has already been performed in
\cite{Cli}.  We will simply quote the result of this study here, the
reader is referred to \cite{Cli} for details of the derivation.

Writing the perturbed line-element as
\begin{eqnarray*}
ds^{2}=&-r^{2\delta \frac{(1+2\delta )}{(1-\delta )}}(1+V(r))dt^{2}\\&+\frac{%
(1-2\delta +4\delta ^{2})(1-2\delta -2\delta ^{2})}{(1-\delta )^{2}}%
(1+W(r))dr^{2}+r^{2} d\Omega^2,
\end{eqnarray*}
substituting this into the field equations and linearising them to
first order in $V$ and $W$ allows the general solution to first order
in perturbations to be found as
\begin{eqnarray*} 
V(r) &= c_1 V_1(r)+c_2 V_2(r)+ c_3 V_3(r) +\rm{constant} \\
W(r) &= -c_1 V_1(r)+c_2 W_2(r)+c_3 W_3(r)
\end{eqnarray*}
where 
\begin{eqnarray*}
V_1 &=-r^{-\frac{(1-2 \delta+4 \delta^2)}{(1-\delta)}} \\
V_2 &= \frac{ (1+2 \delta) r^{-\frac{(1-2 \delta+4 \delta^2)}{2 (1-\delta)}}%
}{2 (2-3 \delta+12 \delta^2+16 \delta^3)} \left( (1+2 \delta)^2 \sin (A \log
r) + 2 A (1-\delta) \cos(A \log r) \right) \\
W_2 &= r^{-\frac{(1-2 \delta+4 \delta^2)}{2 (1-\delta)}} \sin (A \log r) \\
V_3 &= \frac{ (1+2 \delta) \left( (1+2 \delta)^2 \cos (A \log
r) - 2 A (1-\delta) \sin(A \log r) \right) r^{-\frac{(1-2 \delta+4 \delta^2)}{2 (1-\delta)}}
}{2 (2-3 \delta+12 \delta^2+16 \delta^3)} \\
W_3 &= r^{-\frac{(1-2 \delta+4 \delta^2)}{2 (1-\delta)}} \cos (A \log r)
\end{eqnarray*}
and 
\begin{equation*}
A= -\frac{\sqrt{7-28 \delta+36 \delta^2-16 \delta^3-80 \delta^4}}{2
(1-\delta)}.
\end{equation*}
The extra constant in $V$ can be absorbed into the definition of the
time coordinate.  This procedure of solving the linearised field
equations does not ensure that the solution obtained will be the
linearisation of the general solution to the full non-linear field
equations, but it is encouraging to note that the $c_2$ mode
corresponds to the linearisation of the exact solution (\ref{Chan}).

\subsection{Solution 2}

The corresponding perturbative analysis about the background given by
(\ref{Fon}) will now be performed.  Writing the line-element as
\begin{equation}
\label{lin5}
ds^2=-(1+P(r)) dt^2+b^2(t) (1+Q(r)) (dr^2+r^2 d\Omega^2)
\end{equation}
allows the vacuum field equations to be linearised in $P$ and $Q$.
These linearised field equations are given in the appendix, and are
solved to give the solutions
\begin{eqnarray*}
P &= -\frac{c_4}{r} +\frac{2 c_5 (1-6 \delta+4 \delta^2+4
  \delta^3)}{(5-14 \delta-12 \delta^2)} r^2+\rm{constant}\\
Q &= \frac{(1-2 \delta) c_4}{r} +\delta c_5 r^2 +\rm{constant}
\end{eqnarray*}
where $c_4$ and $c_5$ are constants and the two other constant terms
in $P$ and $Q$ are independent of each other and can be absorbed into
$t$ and $s$ by redefinitions.  Again, it can be seen that one of the
modes, $c_4$, corresponds to the linearised version of the exact solution,
(\ref{Fon}).

It can immediately be seen that the form of these linearised solutions
are quite different to those obtained by expanding around the
background of solution 1, (\ref{Chan}).  Whist the expansion about
(\ref{Chan}) produces damped oscillatory modes, as well as the mode
corresponding to the linearised exact solution, the expansion about
(\ref{Fon}) produces more familiar looking terms proportional to
$r^2$.  Aside from the different form of these extra modes there is
also a noticeable difference in the terms corresponding to the
linearised exact solutions, which both go as $r^{-1}$ in the limit
$\delta \rightarrow 0$, but behave differently from each other when
$\delta \neq 0$.  This shows explicitly the differences
that can arise when linearising about different backgrounds.  Not only
are there extra modes which can take different functional forms,
but even the modes which reduce to the Schwarzschild limit as $\delta
\rightarrow 0$ are different, depending on the background that has been chosen.  

We will now proceed to calculate the equations of motion of particles
following geodesics of these space-times, to Post-Newtonian order.  It will be shown that not
only are the terms due to the linear perturbations different (as
should be expected as the perturbations themselves have been shown to
be  background dependant) but that the background itself also
contributes extra terms to the post-Newtonian equations of motion.

\section{Equations of motion}

We will now proceed to calculate the geodesics of the two space-times,
(\ref{Chan}) and (\ref{Fon}).  In doing this we will neglect the
contributions from the oscillating modes to (\ref{Chan}) and the
contribution of the $r^2$ mode to (\ref{Fon}) so that we only take
into account the modes which go as $r^{-1}$, in the limit $\delta
\rightarrow 0$.  These are the modes corresponding to the
linearisation of the exact solutions (\ref{Chan}) and (\ref{Fon}).  In
performing this computation we will transform the
solution (\ref{Chan}) into isotropic coordinates (details of this
transformation are given in \cite{Cli}).

The geodesic equation can be written, as usual, in the form
\begin{equation*}
\frac{d^{2}x^{\mu }}{d\lambda ^{2}}+\Gamma _{\;ij}^{\mu }\frac{dx^{i}}{%
d\lambda }\frac{dx^{j}}{d\lambda }=0,
\end{equation*}
where $\lambda $ can be taken as proper time for a time-like geodesic, or as
an affine parameter for a null geodesic. In terms of coordinate time this
can be re-written as
\begin{equation}
\label{geo9}
\frac{d^{2}x^{\mu }}{dt^{2}}+\left( \Gamma _{\;ij}^{\mu }-\Gamma _{\;ij}^{0}%
\frac{dx^{\mu }}{dt}\right) \frac{dx^{i}}{dt}\frac{dx^{j}}{dt}=0.
\end{equation}
Substituting the linearised solutions into this equation will then give the
equations of motion for test particles in these space-times.

\subsection{Solution 1}

In isotropic coordinates the linearised version of (\ref{Chan}) can be
written as
\begin{equation*}
ds^2=-A_3(r) dt^2 +B_3(r) (dr^2+r^2 d \Omega^2)
\end{equation*}
where
\begin{eqnarray*}
A_3(r) &= r^{\frac{2 \delta (1+2 \delta)}{n}} \left(1-\frac{(1-2
  \delta) c_6}{r^m} \right)\\
B_3(r) &= r^{-2+\frac{2 (1-\delta)}{n}}
      \left(1+\frac{c_6}{r^m} \right)
\end{eqnarray*}
and
\begin{eqnarray*}
n^2 &= (1-2 \delta-2 \delta^2)(1-2 \delta+4 \delta^2)\\
m^2 &= \frac{(1-2 \delta+4 \delta^2)}{(1-2
      \delta -2\delta^2)}
\end{eqnarray*}
and $c_6$ is constant.  Substituting this metric into the geodesic
equation (\ref{geo9}) gives, to post-Newtonian order,
\begin{eqnarray}
\label{eqnm}
\frac{d^2 {\bf{x}}}{dt^2} =&-\frac{\delta (1+2 \delta)}{n
  r^{\frac{2}{m}-1}} {\bf{e}}_r -\frac{(1-8 \delta+4 \delta^2) c_6}{2 n r^{\frac{3
  (1-2 \delta)}{n}-1}} {\bf{e}}_r+\frac{(1-8 \delta+4 \delta^2) c_6^2}{2 n r^{\frac{4
  (1-\delta)^2}{n}-1}} {\bf{e}}_r\\ 
&+\left( \frac{(1-\delta-n)}{n r}-\frac{m c_6}{2 r^{1+m}}\right) \left(
  \frac{d {\bf{x}}}{dt} \right)^2 {\bf{e}}_r \nonumber
\\&+\left( \frac{2 (m-1)}{m r}+\frac{2 (1-\delta) m c_6}{r^{1+m}} \right)
  {\bf{e}}_r \cdot \frac{d{\bf{x}}}{dt} \frac{d{\bf{x}}}{dt}, \nonumber
\end{eqnarray}
which to first order in $\delta$ is
\begin{equation}
\label{eqnm2}
\frac{d^{2}{\bf{x}}}{dt^{2}}=-\frac{Gm}{r^{2}}\left( 1+\left\vert \frac{d%
{\bf{x}}}{dt}\right\vert ^{2}\right) {\bf{e}_{r}}+4\frac{G^{2}m^{2}}{%
r^{3}}{\bf{e}_{r}}+4\frac{Gm}{r^{2}}{\bf{e}_{r}}\cdot \frac{d{\bf{x}}}{%
dt}\frac{d{\bf{x}}}{dt} -\frac{\delta }{r} {\bf{e}}_r
\end{equation}
where the Newtonian limit has been used to set $c_6$.  It can be seen
that these equations of motion are modified from their usual form in
GR, both in the pre-multiplicative factors of the terms with GR
counterparts as well as modifications to the powers of $r$ and the
existence of entirely new terms which vanish in the limit $\delta
\rightarrow 0$.  The first order corrections due to small, but non-zero,
$\delta$ are primarily due to the new term arising in (\ref{eqnm2}).
This term corresponds to a new force which drops off as $r^{-1}$ and
is the term which was used in \cite{Cli} to impose upon $\delta$ the
tight constraint
\begin{equation}
\delta =2.7\pm 4.5\times 10^{-19}
\end{equation}
from observations of the perihelion precession of Mercury \cite{Sha76}.

\subsection{Solution 2}

Solution 2 has already been given in isotropic coordinates in equation
(\ref{Fon}), which on substitution into (\ref{geo9}) gives to
post-Newtonian order the equation of motion
\begin{eqnarray}
\label{eqnn}
\frac{d^2{\bf{x}}}{d\tau^2} = &-\frac{Gm}{r^2} {\bf{e}}_r+4
(1-\delta) \frac{G^2m^2}{r^3} {\bf{e}}_r -(1-2 \delta) \frac{G
  m}{r^2} \left( \frac{d{\bf{x}}}{d \tau} \right)^2 \\&+4 (1-\delta)
\frac{G m}{r^2} {\bf{e}}_r \cdot \frac{d{\bf{x}}}{d \tau}
\frac{d{\bf{x}}}{d \tau} -\mathcal{H} \frac{d{\bf{x}}}{d \tau}
\left(1- \left(\frac{d{\bf{x}}}{d \tau} \right)^2 \right) \nonumber
\end{eqnarray}
where
\begin{equation*}
\mathcal{H} \equiv \frac{1}{a} \frac{da}{d \tau},
\end{equation*}
the conformal time coordinate $\tau$ is defined by $dt \equiv a d\tau$
and $c_4$ has been set by the appropriate Newtonian limit.  The equations
of motion for this solution are considerably simpler than those of
solution 1, but still differ from those of GR in significant ways.
All of the terms except the last in equation (\ref{eqnn}) have GR
counterparts and the powers of $r$ in these terms are all the same as
in the general relativistic case.  The premultiplicative factors of these
terms are, however, modified and can be described adequately within
the frame-work of the parameterised post-Newtonian approach \cite{will} by assigning
\begin{equation*}
\beta = 1 \qquad {and} \qquad \gamma = 1-2 \delta.
\end{equation*}
By making this identification the constraints on $\gamma$ from
observations of the Shapiro time delay of radio signals from the
Cassini space probe \cite{Bert} can be used to
impose upon $\delta$ the constraint
\begin{equation}
\delta = -1.1 \pm 1.2 \times 10^{-5}.
\end{equation}
As well as the usual effects associated with $\gamma-1$ being non-zero
there are extra effects in this space-time due to the last term in
(\ref{eqnn}).  This term is proportional to the velocity of the
test-particle (when $v << c$) and is zero for photons.  For this reason we
identify it as a friction term, with the friction coefficient being
given by $\mathcal{H}$.  This `friction' is a purely gravitational
effect and is not due to any non-gravitational interaction of test particles with
any other matter.

\section{Conclusions}

We have considered here the problem of finding spherically symmetric
vacuum solutions of fourth-order theories of gravity.  In these
theories the Einstein-Hilbert term may not be the dominating one at low
curvatures in the generating
Lagrangian and the validity of taking a Minkowski or de-Sitter background to
expand about cannot be assumed so readily.  As well as this, Birkhoff's
theorem is not valid in fourth-order theories, which allows for a much
larger manifold of solutions, when spherical symmetry is imposed.
These features allow for the interesting possibility of having a
choice of backgrounds to expand about, both static and non-static,
which can vary considerably in form from their general relativistic
counterparts.

We have presented two exact solutions which are
spherically symmetric and reduce to the Schwarzschild solution in
the limit of the fourth-order theory reducing to general relativity.
These solutions have very different backgrounds ($r \rightarrow \infty$
limits) and gravitational physics in each of them is correspondingly
different.  The first of these solutions is static and has a
non-trivial dependence on the radial coordinate $r$, as $r \rightarrow
\infty$.  The second is non-static and has an $r$ independent form in
the limit $r \rightarrow \infty$.  In some sense these solutions can be
considered as being extremes of the general solution; in the first
case the strong curvature at large distances from the centre of
symmetry is only dependent on $r$ and
in the second the curvature at large distances is only a function of
$t$ (in the limit $r \rightarrow \infty$).

Using these exact solutions we have shown how different choices of background lead to
very different solutions of the linearised field equations, when
spherically symmetric perturbations are introduced.  By taking the
perturbation modes which reduce to something approximating Newtonian
gravity, in the appropriate limit, we have derived the equations of
motion for test particles following geodesics of these space times.
The equations of motion are found to be strongly dependent on the
choice of background.  In both cases the form of the terms that have counterparts in GR are
dependent on the background.  As well as this dependence there are
additional terms due to the background itself which have no
counterparts in the general relativistic solutions.  For the static
case these extra terms look like a fifth force which drops off as
$\sim r^{-1}$ and for the time dependent case they look like a
friction term $\propto v$, when $v << c$.

These considerations show explicitly that there are a number of
different backgrounds about which one may choose to perform a
perturbative expansion, in fourth-order theories, and that these backgrounds
can display behaviour which is not permitted in general relativity.
Moreover, it has been shown that the choice of background is highly non-trivial when performing
a perturbative expansion.  The form of the perturbations are
background dependent and, correspondingly, so are the geodesics of the
perturbed space-times.

The existence of multiple backgrounds, all of which appear to be valid
solutions of the field equations, prompts the question: which is the
appropriate solution for physical situations?  We will not attempt to
address this question fully here, but will offer a few suggestions and
leave a more comprehensive analysis for a future study.  Firstly, a physically
relevant background solution should be stable to perturbations.
Imposing the extra symmetry of time-independence it was shown in \cite{Cli} that the
$r \rightarrow \infty$ limit of solution (\ref{Chan}) is the attractor
of the general spherically symmetric and time independent solution, in this limit.  It was also
shown in reference \cite{Cli} that the $r \rightarrow \infty$ limit of
solution (\ref{Fon}) is the attractor of the general homogeneous and
isotropic vacuum solution in the limit $t \rightarrow \infty$ (at
least for some range of $\delta$).  In this sense both of these
solutions can be considered as stable, but this does not allow us to
differentiate between the two of them which is the physically more relevant.  In order to establish the
appropriate solution for any realistic physical situation it will be
necessary to apply boundary conditions to either or both of an
interior solution for an energy-momentum distribution and/or the
relevant cosmological solution.  Whilst a number of cosmological
solutions are known \cite{Cli, Bar06, further, Clif} this problem is
made more difficult by a lack of exact solutions which may be valid in
the interior of, for example, a star.  A more comprehensive analysis
is also required in order to establish the effect of allowing the
perturbations themselves to be time dependent.

\section{Appendix}

Substituting the perturbed line-element (\ref{lin5}) into the vacuum
field equations (\ref{fieldq}) allows the following set of coupled,
linearised differential equations to be found.  

The $t-t$ equation is
\begin{eqnarray*}
\fl \qquad \qquad
6 \delta (1-2 \delta -2 \delta^2) (5-14 \delta -12 \delta^2) \nabla P
\\-12 (1-2 \delta -2 \delta^2) (1-6 \delta +4 \delta^2+4 \delta^3)
\nabla Q \\+(1+2 \delta) (1-\delta)^2 t^{\frac{2 (1-2 \delta-2
    \delta^2)}{(1-\delta)}} \nabla \psi =0
\end{eqnarray*}
the $r-r$ equation is
\begin{eqnarray*}
\fl \qquad \qquad
(2- 17 \delta +18 \delta^2 +52 \delta^3 +8 \delta^4) P'' -2 (1
  -\delta +3 \delta^2 -34 \delta^3 -32 \delta^4) \frac{P'}{r} \\+2 (2 -8
  \delta -9 \delta^2 +16 \delta^3 +8 \delta^4) Q''+ 2 (1 -4 \delta -15
  \delta^2 +20 \delta^3 +16 \delta^4) \frac{Q'}{r} \\+3
  (1-\delta)^2 t^{\frac{2 (1-2 \delta-2 \delta^2)}{(1-\delta)}} \left(
  \frac{\psi''}{2 (1+2 \delta)} +\frac{\psi'}{3 r} \right)=0
\end{eqnarray*}
and the $t-r$ equation is
\begin{equation*}
\psi'=0
\end{equation*}
where 
\begin{equation*}
\psi=\nabla P + 2 \nabla Q,
\end{equation*}
primes here denote differentiation with respect to $r$ and $\nabla$
is the Laplacian on the three dimensional subspace.  The
$\theta-\theta$ and $\phi-\phi$ equations are then linear combinations
of the above equations.

\ack

The author would like to acknowledge John Barrow for helpful
discussions and the PPARC for financial support.


\section*{References}

\begin{thebibliography}{100}

\bibitem{Pech} E. Pechlaner and R. Sexl, Comm. Math. Phys. \textbf{2},
  165 (1966).

\bibitem{Ste77} K. S. Stelle, Phys. Rev. D \textbf{16}, 953 (1977).

\bibitem{Cli} T. Clifton and J. D. Barrow,
  Phys. Rev. D \textbf{72}, 103005 (2005).

\bibitem{Bar06} J. D. Barrow and T. Clifton, Class. Quant. Grav. 
\textbf{23}, L1 (2006).

\bibitem{further} T. Clifton and J. D. Barrow,
  Class. Quant. Grav. \textbf{23}, 2951 (2006).

\bibitem{Car04} S. Carloni, P. K. S. Dunsby, S. Capoziello and
  A. Troisi, Class. Quant. Grav. \textbf{22}, 4839 (2005).

\bibitem{Leach} J. A. Leach, S. Carloni, P. K. S. Dunsby,
  Class. Quant. Grav. \textbf{23}, 4915 (2006).

\bibitem{Schmidt} H.-J. Schmidt, Astron. Nachr. \textbf{311}, 165 (1990).

\bibitem{Clif} T. Clifton and J. D. Barrow, Phys. Rev. D \textbf{72},
  123003 (2005).

\bibitem{RRn} S. Capozziello, F. Occhionero and L. Amendola,
  Int. J. Math. Phys. D \textbf{1}, 615 (1993).

\bibitem{Car05} S. M. Carroll, A. De Felice, V. Duvvuri, D. A. Easson, M.
Trodden and M. S. Turner, Phys. Rev. D \textbf{71}, 063513 (2005).

\bibitem{Cha95} K. C. K. Chan, J. H. Horne and R. B. Mann, Nucl. Phys. 
\textbf{B447}, 441 (1995).

\bibitem{Fonarev} O. A. Fonarev, Class. Quant. Grav. \textbf{12}, 1739 (1995).

\bibitem{Sha76} I. I. Shapiro, C. C. Counselman III and R. W. King, Phys.
Rev. Lett. \textbf{36}, 555 (1976).

\bibitem{will} C. M. Will, \textit{Theory and Experiment in Gravitational
Physics} (rev. edn.), Cambridge University Press, Cambridge (1993).

\bibitem{Bert} B. Bertotti, L. Iess and P. Tortora, Nature \textbf{425}, 374
(2003).

\end{thebibliography}
\end{document}